\documentstyle[prl,aps,multicol,epsf]{revtex}
\begin{document}
\newcommand{\He}{{H_{\rm ex}}}
\newcommand{\et}{{\it et al.}, }
\newcommand{\bi}{\bibitem}
\draft

\title{Two-Phase Region of the Vortex-Solid Melting Transition:  
3D~$XY$ Theory}
\author{Mark Friesen and Paul Muzikar}
\address{Department of Physics, Purdue University, West Lafayette,
Indiana 47907-1396}
\date{\today }
\maketitle
\begin{abstract}
In clean enough samples of the high-$T_c$ oxide materials, the
phase transition into the superconducting state occurs along a
first order line in the $H$-$T$ plane.  This means that a two-phase
region occurs in the $B$-$T$ plane, in which the liquid and solid
vortex phases coexist.  We discuss the thermodynamics of this
two-phase region, developing formulae relating experimental
quantities of interest.  We then apply the 3D~$XY$ scaling theory
to the problem, obtaining detailed predictions for the boundaries
of the coexistence region.  
By using published data, we are able to predict the 
width of the two-phase region, and determine the physical 
parameters involved in the 3D~$XY$ description.
\end{abstract}
\begin{multicols}{2}

The phase transition into the superconducting state in the high-$T_c$
oxide materials is a topic of great current interest.  
It is well known that the
mean-field description of the superconducting transition 
must be modified to incorporate strong fluctuation effects.
In the presence of magnetic fields, 
the resulting transition has been described as the melting
of a vortex-solid into a vortex-liquid, and is
predicted to be first order in very clean samples \cite{brezin}.
This prediction is borne out both in experiments 
\cite{safar,liang,welp,zeldov} and simulations \cite{hetzel}.  
Important thermodynamic support for the first order scenario has been
provided by Schilling {\it et al.}, who verify that the Clausius-Clapeyron
equation is satisfied \cite{schilling}.

In contrast, the zero field transition
($H=0,T=T_c$) is continuous, and belongs to the universality class
of the 3D~$XY$ model \cite{note0}.  The discontinuities of the entropy
density ($\Delta s$) and magnetization ($\Delta M$) therefore both go to 
zero approaching this critical point.  The observed range of the critical
fluctuations extends over large portions of the phase diagram in both
clean \cite{liang} and disordered \cite{salamon} samples.  The 3D~$XY$
theory provides a rigorous framework for
analyzing the first order transition, and thus places 
strict constraints on its interpretation.  Experimental data
supporting this description have been obtained by Liang
{\it et al.}\cite{liang}.

As illustrated in Fig.~1, the first order transition occurs along a
line $H_m(T)$ in the $H$-$T$ plane.  
However, in the $B$-$T$ plane, 
the first order line becomes an area of two-phase coexistence.
This area is analogous to the region of liquid-solid coexistence in the
$T$-$\rho$ (temperature-density) plane of a substance such as water.  
When
$B$ (the average magnetic field) and $T$ are such that the superconductor
is in the two-phase region, the system separates into two phases, each
with its own value of $B$.  It is a point of general and fundamental 
interest to understand the distinction between the $H$-$T$ and $B$-$T$ 
planes; 
as we shall explain,
some nontrivial signatures of the first order transition can be inferred by 
keeping this distinction in mind.

The purpose of this paper is to use 3D~$XY$ scaling to discuss the
two-phase region.  We start by giving a general
thermodynamic analysis, emphasizing the $\He$-$T$ plane,
where $\He$ is the externally applied magnetic field \cite{dodgson}.  
This allows us to 
obtain relations between the different experimental quantities of interest,
thus providing important consistency checks for experimental
investigations.  We then apply the 3D~$XY$ scaling theory to make more
specific predictions concerning the coexistence region.  Background effects 
are carefully included, in terms of the background inverse permeability 
$\Omega$.  We evaluate the important parameters of the scaling 
theory using published data; this allows us to 
make definite predictions about the size of the coexistence region for 
several systems.  

{\it General Theory.}
We consider the coexistence of two phases:  the 
vortex-liquid (L) and the vortex-solid (S).  (These names are used only 
for convenience; the particular nature of the two phases plays no
role in the analysis.)  We assume that each phase is
characterized by a distinct free energy density, denoted as
$f_L(B,T)$ and $f_S(B,T)$.  The conjugate fields $H_\alpha (B,T)$ are defined
in the usual way:
\begin{equation}
H_\alpha (B,T)=4\pi \partial f_\alpha /\partial B ,\label{eq:Hdef}
\end{equation}
where $\alpha = L$ or $S$.
The curves bounding the two-phase region in the $B$-$T$ plane
are denoted by $B_L(T)$ and $B_S(T)$, as in Fig.~1.  
The fundamental equations
describing coexistence along the line $H_m(T)$ are then
\begin{eqnarray}
& & H_L(B_L(T),T)=H_S(B_S(T),T) \equiv  H_m(T) ,\label{eq:Heq} \\
& & f_S(B_S(T),T)-f_L(B_L(T),T) \nonumber \\
& & \mbox{} \quad \quad = H_m(T)[B_S(T)-B_L(T)]/4\pi 
.\label{eq:geq}
\end{eqnarray}
The first equation requires that $H$ be the same in the two phases at a
given temperature, while the second equation
requires that the magnetic Gibbs free energies be the same
at that temperature.  Eqs.~(\ref{eq:Heq}) and (\ref{eq:geq}) 
can be solved simultaneously to
determine the two unknowns, $B_S(T)$ and $B_L(T)$.

For most experimental situations, it is the external field $\He$
which is directly under control.  For an ellipsoidal sample with 
demagnetizing factor $n$,  $\He$ is related to $B$ and $H$ by
\begin{equation}
\He = nB + (1-n)H = B + 4\pi (n-1)M ,\label{eq:ellipse}
\end{equation}
when the field is applied along a symmetry axis of the 
ellipsoid \cite{note1}.

\vspace*{-25truemm}

\begin{figure}
\epsfxsize=3.1truein
\centerline{\epsffile{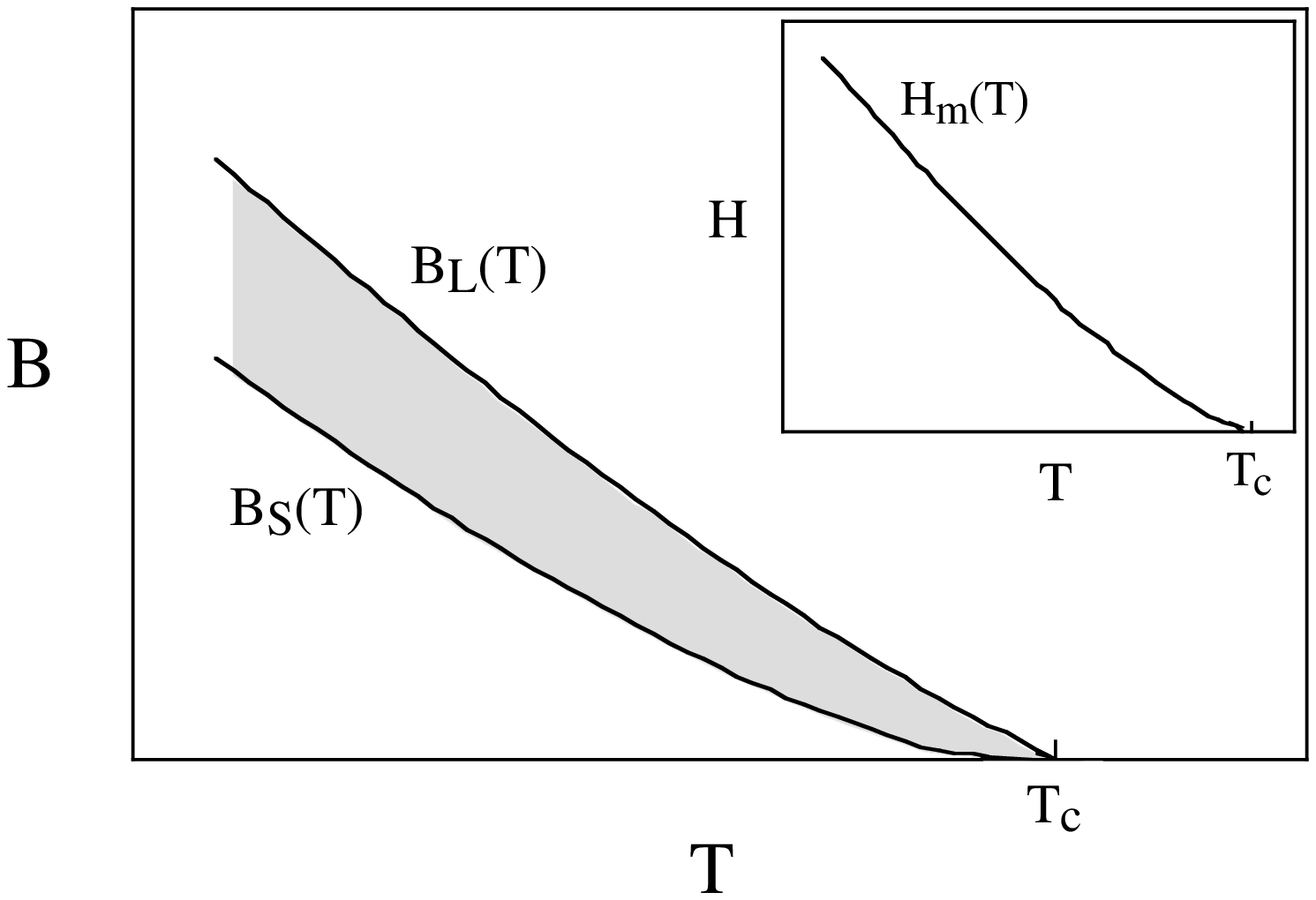}}
\end{figure}
{\small \noindent \mbox{} \hskip.25truecm FIG.~1.
First order melting phase diagram in the $B$-$T$ plane,
showing the two-phase region (shaded).  The coexistence boundaries are
given as $B_S(T)$ and $B_L(T)$.  Inset:  in the $H$-$T$ plane,
coexistence occurs along the line $H_m(T)$.}

\vspace{5truemm}

We will now consider the two most common experimental procedures for 
collecting data:  
(i) varying $H_{ex}$ while keeping $T$ fixed,
(ii) varying $T$ while keeping $H_{ex}$ fixed.  
It will be seen that the distinction between these two cases is nontrivial.

The first results are evident from Eqs.~(\ref{eq:Heq}) and 
(\ref{eq:ellipse}) \cite{farrell}:
\begin{equation}
\Delta \He |_T= n\Delta B |_T = 4\pi n\Delta M|_T. \label{eq:constT}
\end{equation}
Here, the operator $\Delta$ is defined as the value measured on the liquid 
side minus the value measured on the solid side of the 
transition.  The notation of Eq.~(\ref{eq:constT}) also
signifies that the difference is measured at constant $T$ in this case.
Note that the discontinuity $\Delta B|_T$ occurs because fields in the
range $B_S(T)<B<B_L(T)$ are not stable.  However, the discontinuity 
$\Delta \He$ reflects only the boundaries of the two-phase region;
there are no unattainable values of $\He$.

We now discuss the constant-$\He$ path across the coexistence region.
Eq.~(\ref{eq:ellipse}) shows that a 
discontinuity $\Delta B|_\He$ causes corresponding discontinuities in 
$\Delta H|_\He$ and $\Delta M|_\He$.  According to Eq.~(\ref{eq:Heq}), if 
$H$ changes values in the coexistence region, then it must follow the line 
$H_m(T)$.
Since the discontinuities $\Delta B$ and $\Delta H$
are much smaller than $B$ and $H$ respectively (except when 
$T\simeq T_c$), we see that
$\Delta H|_\He /\Delta T|_\He \simeq \partial H_m/\partial T$.  The 
following relations are thus obtained at constant $\He$:
\begin{eqnarray}
\Delta B|_\He & = & (1-1/n) \Delta H|_\He = 
4\pi (1-n) \Delta M|_\He \nonumber \\
& \simeq & (1-1/n)
(\partial H_m /\partial T) \Delta T|_\He .\label{eq:constHe}
\end{eqnarray}
In the final equation, $H$ may be replaced by $\He$ 
using the approximate relation 
$[1+(n-1)(1-\Omega )]H\simeq \Omega \He$, where the inverse background 
permeability $\Omega$ is defined 
below, and satisfies $\Omega \simeq 1$.  Corrections to this 
approximation are of order $M/\He$, and are therefore extremely small 
for typical YBa$_2$Cu$_3$O$_{7-\delta}$ (YBCO) samples, but become more 
noticeable in Bi$_2$Sr$_2$CaCu$_2$O$_8$ (BSCCO).

Equations~(\ref{eq:constT}) and (\ref{eq:constHe}) are related through the 
following geometrical statement:
\begin{equation}
\Delta \He |_T/\Delta T|_\He 
\simeq -\partial H_{{\rm ex}m}/\partial T.\label{eq:nogeom}
\end{equation}
This equation is important in its own right, since it is not affected by the 
sample geometry.  

Using the above results,
we may obtain various relations of interest.  For example, we find
$[1+(n-1)(1-\Omega )]\Delta M|_\He \simeq \Omega \Delta M|_T$, with 
corrections of order $M/\He$.  When $\Omega =1$, this equation reduces 
to $\Delta M|_\He \simeq \Delta M|_T$, which is independent of sample 
geometry.  Similarly for $\Omega =1$, we find 
$\Delta (B-\He )|_T \simeq \Delta B|_\He$, which is also 
geometry-independent.
Finally, we state the Clausius-Clapeyron equation appropriate for this 
transition:
\begin{equation}
4\pi \Delta s|_T/\Delta B|_T =-\partial H_m/\partial T.\label{eq:cc}
\end{equation}
Equations (\ref{eq:constT})-(\ref{eq:cc}) 
form a set of thermodynamic relations between 
quantities of interest for the first order transition.

{\it 3D~$XY$ Scaling Theory.}  The main features of 3D~$XY$ scaling 
can be derived from a single ansatz for the free energy density
\cite{ffh,schneider,xytheory}.  
The new assumption for the coexistence region 
is that the liquid and solid free energies densities independently satisfy this 
scaling.  This means that we may write the free energies in the following
form, using $t\equiv |T-T_c|/T_c$ \cite{xytheory}:
\begin{eqnarray*}
f_L(B,T) &=& f_{kL}t^{3\nu}\phi_L(Bt^{-
2\nu}/H_{kL})+f_0(T)+\Omega B^2/8\pi  \\
f_S(B,T) &=& f_{kS}t^{3\nu}\phi_S(Bt^{-
2\nu}/H_{kS})+f_0(T)+\Omega B^2/8\pi . \\
\end{eqnarray*}
Note the following points:

(1) The exponent $\nu \simeq 0.67$ is the correlation length exponent
for the 3D~$XY$ model.  The central assumption of the scaling ansatz
is that the magnetic field $B$ scales as an inverse length 
squared \cite{ffh}.  $B$ and $T$ then enter the scaling function only
in the combination $Bt^{-2\nu}$.

(2)  The parameters $f_{kL}$, $f_{kS}$, $H_{kL}$, and $H_{kS}$ are
material dependent and thus nonuniversal.  $f_k$'s have dimensions of
free energy density, while $H_k$'s have dimensions of magnetic field.  
Below, we show that $f_{kL}=f_{kS}$.  

(3)  The term $f_0(T)+\Omega B^2/8\pi$ approximates the smooth 
background contribution to the free energy density near the zero field
transition.  The quantities $f_0(T)$ and $\Omega (T)$ are nonuniversal
and contain no singularities near $T\simeq T_c$.
The inverse background permeability $\Omega (T)$ is only weakly 
temperature
dependent in many cases, and satisfies $\Omega \simeq 1$. 
We assume that the background terms have the same form in both the
solid and liquid phases.  

(4)  The two scaling functions $\phi_S$ and $\phi_L$ are distinct and
universal.  However, the two phases should become 
indistinguishable when $B=0$.  
Thus, we expect the following equality to hold:
\begin{equation}
f_{kL}\phi_L(0)=f_{kS}\phi_S(0) .\label{eq:zeroB}
\end{equation}

(5)  The conjugate fields are obtained from Eq.~(\ref{eq:Hdef}):
\begin{displaymath}
H_\alpha (B,T) = \Omega B
+(4\pi f_{k\alpha}/H_{k\alpha})t^\nu \phi'_\alpha 
(Bt^{-2\nu}/H_{k\alpha}) 
\end{displaymath}
where $\alpha =L$ or $S$.
We now use Eqs.~(\ref{eq:Heq}) and (\ref{eq:geq}) to determine the
coexistence boundary lines $B_S(T)$ and $B_L(T)$.  This is accomplished
by noting that all terms in the free energy
involving $\phi_\alpha$ explicitly
(the superconductivity terms)
are small compared to the remaining background terms.
(This assumption breaks down very near the zero field transition, where a 
more careful treatment is 
required \cite{xytheory}.)
The following explicit solutions are obtained:
\begin{eqnarray*}
B_L(T) &=& B^*t^{2\nu}+bt^\nu , \\
B_S(T) &=& B^*t^{2\nu}-bt^\nu .
\end{eqnarray*}
According to our assumptions, $B^*\gg b$ here, while the apparent 
negative value of $B_S(T)$ (when $T\simeq T_c$) is unphysical.
The field-like parameters $B^*$ and $b$ are determined from the 
following equations:
\begin{eqnarray*}
& & f_{kS}\phi_S(B^*/H_{kS})=f_{kL}\phi_L(B^*/H_{kL}) ,\\
& & b=(2\pi/\Omega )[ (f_{kS}/H_{kS})\phi'_S(B^*/H_{kS})- \\
& & \mbox{} \quad \quad (f_{kL}/H_{kL})\phi'_L(B^*/H_{kL})] .
\end{eqnarray*}

The width of the coexistence region in the $B$-$T$ plane can now be
computed along the constant-$T$ path:
\begin{equation}
\Delta B|_T \equiv B_L(T)-B_S(T) =2bt^\nu . \label{eq:DB}
\end{equation}
Similarly, along a constant-$B$ path, the temperature width of the
coexistence region is given to leading order by
\begin{equation}
\Delta T|_B=(T_cb/\nu B^*)(B/B^*)^{(1/2\nu )-1/2} . \label{eq:DT}
\end{equation}
Using Eqs.~(\ref{eq:constT})-(\ref{eq:cc}), these discontinuities can be
related to other quantities of interest.  For example, the magnetization
discontinuity $\Delta M|_T$ is found to scale in the same way as the 
magnetization in the 3D~$XY$ theory, as 
first pointed out by Liang {\it et al.}

In the $H$-$T$ plane, the phase transition is given by
\begin{displaymath}
H_m(T) =\Omega B^* t^{2\nu}+
[\Omega b +(4\pi f_{kL}/H_{kL})\phi'_L(B^*/H_{kL})]
t^\nu ,
\end{displaymath}
where the $t^\nu$ term is again much smaller than the $t^{2\nu}$ term.
The coexistence boundaries in the $\He$-$T$ plane can then 
easily be computed using Eq.~(\ref{eq:ellipse}).
The entropy discontinuity can be calculated to leading order from
Eq.~(\ref{eq:cc}).  For simplicity, we give the result in terms of $B$:
\begin{equation}
\Delta s|_T = (bB^*\Omega \nu /\pi T_c )
(B/B^*)^{(3\nu -1)/2\nu} 
.\label{eq:ds}
\end{equation}

Finally, we show that the material parameters $f_{kS}$ and $f_{kL}$ 
must be equal.  Eq.~(\ref{eq:zeroB}) can first be rewritten
as $f_{kL}/f_{kS}=\phi_{kS}(0)/\phi_{kL}(0)$.  Since the scaling 
functions $\phi_{kS}$ and $\phi_{kL}$ are universal, the ratio
$\phi_{kS}(0)/\phi_{kL}(0)$ must also be universal.  $f_{kL}$ and 
$f_{kS}$
are therefore related by a universal proportionality constant, which
can be absorbed into either $\phi_{kS}$ or $\phi_{kL}$.  Therefore
$f_{kL}=f_{kS}$.  

{\it Comparison with Experiments.}  We now use published data to 
obtain preliminary estimates for the 3D~$XY$ field parameters $B^*$ and 
$b$, and thus the width of the coexistence region, for several systems.
Additionally, we check the thermodynamic relations
(\ref{eq:constT})-(\ref{eq:nogeom}).  In order to apply our analysis to 
real physical situations, it is necessary to assume a particular sample 
geometry.  In accordance with the discussion above, we 
approximate the different samples as ellipsoids, inscribed within the given 
dimensions.  However, we note that several recent papers have considered 
the differences between ellipsoidal geometries and more 
realistic plate-like geometries, which are more difficult to treat 
exactly \cite{farrell,zeldov2}.  Although geometrical effects of this type may 
partially account for the inconsistencies discussed below, we believe that 
some of the discrepancies may be too large to be attributable to geometry 
alone.  We also make the approximation that $\Omega = 1$.

The first experiments we shall consider, in very pure, optimally-doped
YBCO single crystals, 
typically involve bulk magnetization or specific heat data obtained at
constant $T$ or $\He$.  In Ref.~\onlinecite{liang}, Liang {\it et al.} 
observe a discontinuity in $M$ which is well described by the 3D~$XY$ 
theory throughout the studied field range, 5-40~kG.  
(For simplicity, the different fields are all specified
here in units of Gauss.)  Using the results above, and
assuming $n\simeq 0.93$ as appropriate for this sample, we find that
$B^*\simeq 1.0\times 10^6$~G and $b\simeq 0.66$~G.  
To take a particular example then, the width of 
the coexistence region should be equal to
$\Delta T|_\He \simeq 4\times 10^{-5}$~K when $\He =40$~kG.  
We have also obtained estimates for $B^*$ and $b$ from two other 
published magnetization studies of YBCO, with results nearly identical to 
those given above \cite{welp,schilling}.  These values may therefore be 
characteristic of optimally doped YBCO \cite{note2}.

According to the estimates given above, the coexistence region may be 
difficult to resolve along 
the temperature axis, due to its narrow width.
However in the experiments of Schilling {\it et al.} \cite{schilling}, 
estimates for $\Delta \He|_T$ and 
$\Delta M|_T$ can be obtained simultaneously, thus allowing 
Eq.~(\ref{eq:constT}) to be tested.  The following results are observed 
when $T=85$~K:
$4\pi \Delta M|_T\simeq 0.2$~G and $\Delta \He|_T\simeq 1500$~G.  
(Similar results are obtained in Ref.~\onlinecite{welp}.  Note that the 
quantity $\Delta T$ studied in Ref.~\onlinecite{schilling} is unrelated to 
the quantity $\Delta T|_\He$ discussed above.)
According to Eq.~(\ref{eq:constT}), the observed value of 
$\Delta \He|_T$ is too large by a factor of 7500.  On the other 
hand, Schilling {\it et al.}\ have shown that the Clausius-Clapeyron 
equation is obeyed in this system.  It seems likely therefore that the former 
discrepancy does not signal the 
failure of the thermodynamic relations; instead, it may only reflect sample 
dependent effects, which may be generically identified as ``broadening."  
Further experiments are required to resolve this issue.  In any event, the 
predicted width of the coexistence region 
in YBCO appears to be much narrower than the typical experimental 
resolution.  Observed jumps in the different measured quantities should 
therefore be treated as true discontinuities.

Bulk measurements of magnetization discontinuities have also been 
performed in pure BSCCO single 
crystals.  (For example, Ref.~\onlinecite{farrell}.)  However, as our 
second example we consider the experiments of 
Zeldov {\it et al.}, which involve small Hall sensors \cite{zeldov}.  
These devices are capable of measuring the local 
field $B$, thus avoiding many problems due to sample 
inhomogeneity.  The following approximate information can be obtained
from Ref.~\onlinecite{zeldov} for $T\simeq 80$~K:
$\Delta B|_\He \simeq 0.4$~G,
$\Delta T|_\He <3$~mK,
$\Delta (B-\He )|_T \simeq 0.4$~G, 
$\Delta \He|_T <0.4$~G, and
$\partial H_{{\rm ex}m} /\partial T \simeq -6$~G/K.  The appropriate
demagnetizing factor for this sample is $n\simeq 0.71$.
Using these estimates, 
the geometry-independent ($\Omega \simeq 1$) relation 
$\Delta (B-\He )|_T \simeq \Delta B|_\He$ is well satisfied. 
The relation $\Delta (B-\He )|_T =-(1-1/n)\Delta \He |_T$
is not quite satisfied; this small discrepancy may be attributable
to geometric effects.  However,
relation (\ref{eq:constHe}) between $\Delta B|_\He$ and
$\Delta T|_\He$ shows that the observed discontinuity
$\Delta T|_\He$ is at least 50 times sharper than predicted.
Similarly, if the estimates given above for $\Delta T|_\He$ and
$\Delta \He |_T$ are taken as equalities, the geometry-independent
result of Eq.~(\ref{eq:nogeom}) is violated by a factor of more than 20.
To improve this situation, we would need to assume that $\Delta \He |_T$ 
was 20 times smaller than
estimated.  Unfortunately, such a modification leads to discrepancies in 
other thermodynamic relations.

The discrepancies observed in the local $B$ measurements of BSCCO
are of a different nature than those in YBCO.  In the latter case, 
the observed width of 
the two-phase region was much too large, a fact which can be 
attributed to broadening.  In the BSCCO case, the observed two-phase 
region was much too narrow, a result which cannot be explained by any 
obvious thermodynamic 
arguments.  However, it is instructive to continue with the present 
analysis to obtain the 3D~$XY$ parameters appropriate for BSCCO.
According to Ref.~\onlinecite{zeldov}, a reasonable estimate for $B^*$ is 
given by $B^*\simeq 1000$~G.  This result is not affected by the 
inconsistencies described above, and is corroborated by other experiments.  
Assuming the value given above for $\Delta (B-\He )|_T$ to be the most 
reliable estimate of the discontinuities in BSCCO, we then obtain
$b\simeq 2.8$~G.  From Ref.~\onlinecite{farrell} (using low transverse
fields, Fig.~10) we obtain a slightly different result of $b=0.8$~G.

{\it Conclusions.} We conclude by discussing two points.

(1) The results given above for first order melting near the 3D~$XY$ 
critical point differ in several ways from earlier
analyses aimed at fluid-like critical points \cite{rehr}: 
(i) in the 3D~$XY$ scaling 
ansatz, it is the density-like variable ($B$), not the pressure-like
variable ($H$), which appears naturally in the scaling functions;
(ii) in contrast with the fluid system, neither of the scaling variables
$B$ or $H$ are directly related to the superconducting order parameter;
(iii) although a relation like $H_{kS}=H_{kL}$ occurs in
fluid-like scaling theories, the same cannot be proven for the
vortex-solid melting transition in the absence of a more microscopic 
theory.

(2) Our analysis should also apply to computer simulations aimed at 
understanding the first order transition \cite{hetzel}.  Many of these 
simulations are performed by varying the temperature at fixed $B$.  In 
this
case, a two-phase region should be observed, 
with width $\Delta T|_B$.  If the value 
of $B$ is small enough for 3D~$XY$ scaling to be valid, this width should 
be described by Eq.~(\ref{eq:DT}), and the entropy jump by 
Eq.~(\ref{eq:ds}).

In summary, we have presented both general and specific (3D~$XY$) 
theories of
the two-phase region for the vortex-solid melting transition.
We find that in typical YBCO samples, the solid-liquid coexistence region
may be too small to be observed experimentally.  However, based upon 
available estimates for the discontinuity of $B$, the
region should be easily detected in BSCCO
using precision measurements.  The observations of two-phase coexistence 
and thermodynamic self-consistency are both crucial for 
confirming the first order nature of the melting transition.  The recent 
experiment of Schilling {\it et al.}\ provides a check of the self-consistency 
relations; in this case the 
Clausius-Clapeyron equation is well satisfied, although open questions 
remain, regarding other relations.  It is hoped that future experiments may 
address this issue.  Additionally, the scaling of the coexistence
region provides information regarding the most fundamental fluctuations
associated with superconductivity, which are governed by the 3D~$XY$
critical point.

We thank H.\ Nakanishi and S.\ Teitel for helpful discussions.
This work was supported through the Midwest Superconductivity
Consortium (MISCON) DOE Grant No.\ DE-FG02-90ER45427.

%

\end{multicols}
\end{document}